\documentclass[letterpaper, 10 pt, conference]{ieeeconf}  

\IEEEoverridecommandlockouts                              

\usepackage{amsmath,amssymb,color,cite}
\usepackage{graphicx}
\usepackage{subfigure}

\usepackage{latexsym}
\usepackage{algorithm}
\usepackage{algpseudocode}

\usepackage{theorem}

\makeatletter
\algnewcommand{\LineComment}[1]{\Statex \hfill \(\triangleright\) #1}
\makeatother

\usepackage{verbatim}
\usepackage{cite}
\usepackage[english]{babel}

\renewcommand{\theenumi}{(\roman{enumi})}

\usepackage{pgf}
\usepackage{pgfplots,tikz}
\pgfplotsset{compat=1.18} 
\usetikzlibrary{tikzmark}
\usepackage{arydshln}

\usepackage{amsfonts}
\usepackage{amssymb}
\usepackage{amsbsy}
\usepackage{bm}
\usepackage{bbm}
\usepackage{color}
\definecolor{comment}{rgb}{0.0078, 0.5020, 0.0353}

\renewcommand{\algorithmicrequire}{\textbf{Input: }}

\newcommand\oprocendsymbol{\hbox{$\bullet$}}
\newcommand\oprocend{\relax\ifmmode\else\unskip\hfill\fi\oprocendsymbol}

\let\leq\leqslant
\let\geq\geqslant

\newcommand{\R}{\mathbb R}
\newcommand{\N}{\mathbb N}

\newcommand{\Np}{\mathbb N_+}

\newcommand{\calT}{\ensuremath{\mathcal{T}}}

\newcounter{todocounter}
\usepackage[colorinlistoftodos]{todonotes}
\newtheorem{theorem}{Theorem}[section]
\newtheorem{proposition}[theorem]{Proposition}

\newtheorem{corollary}[theorem]{Corollary}

\theoremstyle{remark}
\newtheorem{remark}{Remark}

\theoremstyle{definition}
\newtheorem{assumption}{Assumption}

\newtheorem{problem}{Problem}

\allowdisplaybreaks

\title{Online Data-Driven Adaptive Control for Unknown Linear Time-Varying Systems}

\author{Shenyu Liu, Kaiwen Chen, and Jaap Eising
\thanks{\indent This work was partially supported by the National Science Foundation of China under Grant 62203053, the European Union's Horizon 2020 Research and Innovation Program under Grant 739551 (KIOS Centre of Excellence), and SNF/FW Weave Project 200021E\_20397.}
\thanks{S. Liu is with the School of Automation, Beijing Institute of Technology, Beijing, China. E-mail:	{\tt\small shenyuliu@bit.edu.cn}. K. Chen is with the Department of Electrical and Electronic Engineering, Imperial College London, London, SW7 2AZ, UK. E-mail:	{\tt\small kaiwen.chen16@imperial.ac.uk}. J. Eising is with the Department of Mechanical and Aerospace
	Engineering, University of California, San Diego, CA, USA. E-mail: {\tt\small jeising@ucsd.edu}.}%
}

\begin{document}

\maketitle

\begin{abstract}
	This paper proposes a novel online data-driven adaptive control for unknown linear time-varying systems. Initialized with an empirical feedback gain, the algorithm periodically updates this gain based on the data collected over a short time window before each update. Meanwhile, the stability of the closed-loop system is analyzed in detail, which shows that under some mild assumptions, the proposed online data-driven adaptive control scheme can guarantee practical global exponential stability. Finally, the proposed algorithm is demonstrated by numerical simulations and its performance is compared with other control algorithms for unknown linear time-varying systems.
\end{abstract}

\section{Introduction}
In both classical control and modern control theory, the analysis and design of most controllers rely on the explicit knowledge of the plants. This requirement becomes less practical when the system is complex and highly dimensional. One of the recent research interests in the control community focuses on directly controlling the system by only using the data -- that is, the information of inputs/outputs/states -- while skipping the modelling step. For linear time-invariant (LTI) systems, Willems \textit{et al}.’s \textit{fundamental lemma} \cite{JCW-PR-IM-BLMDM:05} states that if a finite-length input-output trajectory of an LTI system satisfies the so-called \textit{persistence of excitation} condition, then any possible input-output trajectories of this system can be obtained from the aforementioned input-output trajectory. This result was leveraged to avoid system identification in control design, and develop purely data-based methods. For instance, \cite{Coulson2019} developed data-driven model predictive control methods, \cite{CDP-PT:19} deals with classical problems, such as stabilization, \cite{Breschi2021} with data-driven model reference control, and \cite{HJVW-JE-HLT-MKC:20} considers the case where the persistence of excitation condition is not met. Some other related developments are for switched systems (\textit{e.g.}, \cite{MR-CDP-PT:22,Eising2022}), delay systems (\textit{e.g.}, \cite{RuedaEscobedo2022}), and general nonlinear systems (\textit{e.g.}, \cite{TD-MS:20_2,MG-CDP-PT:21}).

A particularly interesting class of system is that of linear time-varying (LTV) systems. These  appear in many real life applications, for instance, due to changes in operating conditions (such as temperature, pressure, \textit{etc}.) and mechanical wear. Moreover, LTV systems can also be obtained by linearizing nonlinear systems around trajectories of time-varying operating points. As a natural extension of the well-established data-driven control theory for LTI systems, data-driven control methods for LTV systems have also attracted much attention in recent years. In \cite{Pang2018}, an optimal control scheme for unknown discrete-time LTV systems is proposed, in which the approximate optimal control is obtained via data-driven off-policy \textit{policy iteration}.
Based on the ensemble of input-state trajectories collected offline, \cite{Nortmann2020} shows a different method of data-driven control of LTV systems. \cite{Verhoek2021} extends the \textit{fundamental lemma} to linear parameter-varying systems and develops a data-driven predictive control scheme for such systems. 
Nevertheless, because a sufficient amount of data is required \textit{a priori} to start the control process, the aforementioned methods cannot be  run completely online. In \cite{Baros2022}, an online data-enabled predictive control is modified from the data-enabled predictive control proposed in \cite{Coulson2019}, and it is claimed to be computationally efficient due to the use of fast Fourier transform and the primal-dual formulation in the algorithm. Nevertheless, stability is not guaranteed theoretically for the proposed controller therein.
To this end, \cite{JM-MS-22} proposes a different data-driven control method by combining matrix inequalities and the matrix S-lemma. This method is technically for linear parameter-varying (LPV) systems, but is applicable to systems with time-varying system matrices. However, that work requires the assumption of a known range of variations.

This paper proposes a novel online data-driven adaptive control (ODDAC) algorithm to stabilize LTV systems. In contrast to the aforementioned methods in the literature, our algorithm can run completely online. Meanwhile, we do not impose the usual assumptions on the knowledge of the system matrices: they do not need to be affine in a time-varying parameter, and they can be unbounded in time. The control gain is periodically updated based on the data collected over a short time window, aiming to stabilize the system up to the time of the next update. The functionality of the ODDAC algorithm is also investigated in this paper via a detailed stability analysis, which shows that under some mild assumptions, the closed-loop system is guaranteed to be practically globally exponentially stable.  

\textit{Notation.} Let $\R$ be the real line, $\N$ be the set of all non-negative integers, and $\Np$ be the set of all positive integers. $0_n$ and $I_n$ denote the zero matrix and the identity matrix in $\R^{n\times n}$, respectively, and the subscript $n$ is omitted if the dimension can be determined according to the context. For any symmetric matrix $M\in\R^{n\times n}$, $M\succ 0$ (resp. $M\succeq 0, M\prec 0$, and $M\preceq0$) means that $M$ is positive definite (resp. positive semi-definite, negative definite and negative semi-definite). For any $n,m\in\Np$ and any vector $x\in\R^n$, $|x|$ denotes the $2$-norm of $x$; for any matrix $N\in\R^{n\times m}$, $\Vert N\Vert$ denotes the induced $2$-norm of $N$. For two sets of matrices $\Sigma_1,\Sigma_2\subset\R^{n\times n}$, denote the Minkowski sum as $\Sigma_1\oplus\Sigma_2:=\{A_1+A_2:A_1\in\Sigma_1,A_2\in\Sigma_2\}$.

The rest of the paper is organized as follows. Section~\ref{sec:formulation} introduces the preliminaries. In section~\ref{sec:modeling}, the mechanism of the ODDAC algorithm and the update law of the feedback gain are explained. In Section~\ref{sec:stability}, the stability properties of the closed-loop system with ODDAC are analyzed. In Section~\ref{sec:simulation}, a numerical example to demonstrate the ODDAC algorithm is presented and the performance is compared to that of other control algorithms for unknown LTV systems. Section~\ref{sec:conclusion} concludes the paper with some discussions on future research directions.

\section{Preliminaries}\label{sec:formulation}
In this section, we will introduce the preliminaries, including the system definition, regularity assumptions, and the problem formulation for control design.

Consider a discrete-time LTV system
\begin{equation}\label{sys}
	x(t+1)=A(t)x(t)+B(t)u(t),
\end{equation}
where $x:\N\to\R^n$ is the state; $u:\N\to\R^m$ is the control input; and $A(t), B(t)$ are time-varying matrices of compatible dimensions. 
We assume that these matrices have a bounded rate of variation; that is, they admit a Lipschitz constant as follows.
\begin{assumption}[Lipschitz matrix trajectories]\label{ass:Lipschitz}
	There exists $L\geq 0$ such that for all $t,s\in\N$,
	\begin{equation}\label{eqn:Lipschitz}
		\left\Vert \begin{bmatrix}
			A(t)-A(s)&B(t)-B(s)
		\end{bmatrix}\right\Vert \leq L|t-s|.
	\end{equation}
\end{assumption}
We are interested in stabilizing the system \eqref{sys} without precisely knowing the matrix trajectories $A(t),B(t)$. Nevertheless, the knowledge of an initial feedback gain $K_0$ is required to start such a process. $K_0$ can either be computed via initially known values of $A(0), B(0)$, or empirically assigned based on \textit{a priori} knowledge of the system. We summarize this condition, together with the controllability assumption as follows.
\begin{assumption}[Controllability and initial gain]\label{ass:controllability}
	The matrix pair $(A(t), B(t))$ is controllable for all $t\in\N$. In addition, for some given $\lambda\in(0,1)$, there exist a known positive definite matrix $P_0\in \R^{n\times n}$ and a known matrix $K_0\in \R^{m\times n}$ such that
	\begin{equation}
		(A(0)+B(0)K_0)^\top P_0 (A(0)+B(0)K_0)\preceq \lambda P_0.
	\end{equation}
\end{assumption}
Under Assumption~\ref{ass:controllability}, $u=K_0x$ is a stabilizing feedback control law for the LTI system obtained by ``freezing'' the LTV system~\eqref{sys} at $t=0$.

The main objective of this paper is to solve the following problem:

\begin{problem}[Data-driven control of an LTV system]\label{prob:1}
	Assume that the time-varying matrices $A(t), B(t)$ in~\eqref{sys} are unknown. Under Assumptions~\ref{ass:Lipschitz}, \ref{ass:controllability}, find a control law which makes the closed-loop system \emph{practically globally exponentially stable} (pGES), namely, there exist positive constants $c_1$, $c_2$, and $c_3$ such that the solution of the trajectories of the closed-loop system satisfy
	\begin{equation}
		|x(t)|\leq c_1e^{-c_2t}|x(0)|+c_3,
	\end{equation}
	for all $t\geq 0$ and $x(0)\in \R^n$.
\end{problem}

\section{Data-driven and adaptation mechanism}\label{sec:modeling}

We propose a novel ODDAC algorithm in order to solve Problem~\ref{prob:1}. In this section, we first introduce the timing of data collection and control gain update. To explain in detail how the control gains are selected based on the data, we convert the LTV system \eqref{sys} into a piece-wise LTI system perturbed by a time-varying term. This allows re-formulating the problem of finding a stabilizing control gain into a linear matrix inequality (LMI) problem, which can be solved efficiently.

\subsection{Periodically updated control law}\label{subsec:sketch}
Due to the time-varying nature of~\eqref{sys}, the initial feedback law $u=K_0x$ may not stabilize the system for all $t\in\N$. To this end, the proposed adaptation mechanism periodically updates the feedback gain at time instants $t^S_i=iT, i\in\Np$, where $T\in \Np$ is the period of gain update. Our goal is to find a feedback gain $K_i\in\R^{m\times n}$ for each $i\in\Np$, such that the control law is ``essentially" $u(t)=K_ix(t)$ for all $t\in\calT_i$. Since the system matrices are unknown, the value of $K_i$ needs to be computed in a way that it only depends on the data over a finite time window prior to time $t^S_i$. To be precise, let $T^W\in\Np$ be the length of time window and denote $t^W_i:=t^S_i-T^W$. In the sequel, we denote $\calT_0:=\{0,1,\cdots,t_1^S-1\}$ and
\begin{align*}
	\calT_i&:=\{t^S_i, t^S_i+1,\cdots, t^S_{i+1}-1\},\\
	\calT^W_i&:=\{t^W_i,t^W_i+1,\cdots,t^S_i-1\},	
\end{align*}
for all $i\in\Np$. Following the earlier discussion, we aim to construct a piece-wise control law:
\begin{equation}\label{true_feedback}
	u(t)=\begin{cases}
		K_ix(t), &t\in\calT_i\backslash\calT^W_{i+1},\\
		K_ix(t)+v(t), &t\in\calT^W_{i+1},
	\end{cases}
\end{equation}
where $v(t)\in\R^m$ is random and uniformly bounded for all $t\in\cup_{i\in\Np}\calT^W_i$. 
The reason for introducing the additional signal $v$ to~\eqref{true_feedback} over $\calT_{i+1}^W$ is that we need the collected data to be ``sufficiently exciting" so that the underlying data-driven problem (Problem~\ref{prob:2} to be discussed later) is feasible. Interested readers are referred to \cite{JCW-PR-IM-BLMDM:05,HJVW:21} for further details.

\begin{figure}[ht]
	\centering
	\includegraphics[width=\columnwidth]{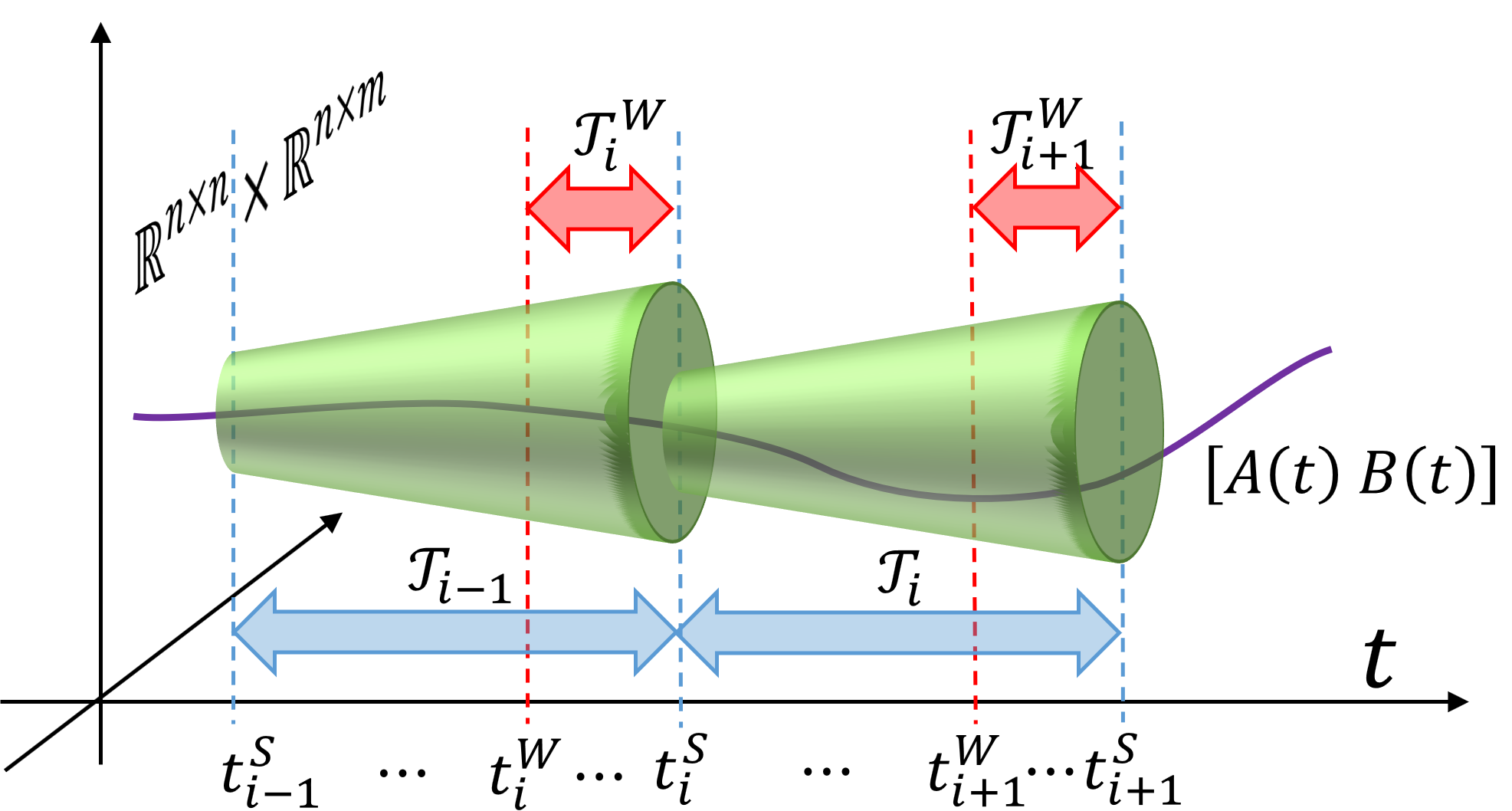}
	\caption{Illustration of the timing of data collection and control gain update. The purple curve represents a ``trajectory'' of $[A(t), B(t)]\in\R^{n\times n}\times \R^{n\times m}$. At each time instant $t_i^S$, the controller estimates a set to which $[A(t),B(t)]$, $t\in\calT_i$, belongs, represented by the green frustum-shaped region. Such estimation is made based on the data collected over the time window $\calT_i^W$.}\label{fig:illustration}
\end{figure}

The timing of data collection and control gain update is illustrated in Fig.~\ref{fig:illustration}. As the system is time-varying, a convergent identification for the matrix trajectories $A(t),B(t)$ is difficult. Instead, based on the data collected over the time window $\calT_i^W$, we aim to first conclude a set to which $A(t^S_i), B(t^S_i)$ belong. The controller further predicts what values $A(t), B(t)$ can reach for $t\in\calT_i$, under Assumption~\ref{ass:Lipschitz}. It then finds a \emph{common} stabilizing feedback gain $K_i$ for the system for all possible values of $A(t), B(t),t\in\calT_i$.

\begin{remark}[Choice of the time window length]
	In principle, one can assign a time-dependent length $T_i^W$ to the $i$th time window, and $T_i^W$ can be selected up to $t_i^S$ (in which case all the history data since $t=0$ is used). The potential advantage of using a variable window length will be studied in future work whereas in this paper we focus on the case in which $T^W$ is fixed and short (in the sense that $T^W<T$).
	By keeping the time window short, the resulting control algorithm is more efficient in both memory use and online computation.
\end{remark}


\subsection{A data-driven model of the LTV system}
		We exploit the idea of the \textit{congelation of variables} method \cite{chen2020adaptive}, which treats the original unknown time-varying system as an unknown time-invariant system (namely, the \textit{congealed} part) perturbed by some time-varying terms. More specifically, the time-varying perturbation terms are defined as $\Delta A_i(t):=A(t)-A(t^S_i)$, $\Delta B_i(t):=B(t)-B(t^S_i)$, for $i\in\Np$. This allows estimating the constant matrices $A(t^S_i), B(t^S_i)$ and attenuating the destabilizing effect of $\Delta A_i(t)$, $\Delta B_i(t)$, separately. To proceed, re-write system~\eqref{sys} as
		\begin{equation}\label{sys:LTI_w_disturbance}
			x(t+1)=A_ix(t)+B_iu(t)+w_i(t),
		\end{equation}
		for $t\in\calT^W_i\cup\calT_i$,	where $A_i:=A(t_i^S)$, $B_i:=B(t_i^S)$ and the ``virtual disturbance" $w_i(t):= \Delta A_i(t)x(t)+\Delta B_i(t)u(t)$.
		We now define the following:
		\begin{subequations}\label{def:X_and_U}
			\begin{align}
				X_i 	&:=\begin{bmatrix}x(t^W_i) & x(t^W_i+1) & \cdots& x(t^S_i-1) \end{bmatrix},\\ 
				X_i^+ 	&:=\begin{bmatrix}x(t^W_i+1) & x(t^W_i+2)& \cdots& x(t^S_i) \end{bmatrix},\\ 
				U_i	&:=\begin{bmatrix}u(t^W_i) & u(t^W_i+1) &\cdots& u(t^S_i-1)\end{bmatrix},\\
				W_i	&:=\begin{bmatrix}w(t^W_i) & w(t^W_i+1) &\cdots& w(t^S_i-1)\end{bmatrix}.
			\end{align}
		\end{subequations}
		Equation \eqref{sys:LTI_w_disturbance} can be re-written into a compact form: 
		\begin{equation}\label{sys:LTI_w_disturbance_stack}
			X_i^+=A_iX_i+B_iU_i+W_i.
		\end{equation}
		Note that if $W_i$ is known and $\begin{bmatrix}
			X_i\\U_i
		\end{bmatrix}$ is full row rank, we can directly compute the values of $A_i, B_i$:
		\begin{equation*}
			\begin{bmatrix}
				A_i& B_i
			\end{bmatrix}=(X^+_i-W_i)\begin{bmatrix}
				X_i\\U_i
			\end{bmatrix}^\dagger ,
		\end{equation*}
		where $(\cdot)^\dagger$ denotes the right inverse.
		However, such computation is infeasible, as $W_i$ depends on unknown $\Delta A_i(t)$ and $\Delta B_i(t)$. Nevertheless, we can employ Assumption~\ref{ass:Lipschitz} and estimate that
		\begin{align*}
			w_i(t)w_i^\top (t)&\preceq |w_i(t)|^2 I \\
			& =\left\vert\begin{bmatrix}
				\Delta A_i(t)&\Delta B_i(t)
			\end{bmatrix}\begin{bmatrix}
				x(t)\\ u(t)
			\end{bmatrix}\right\vert^2 I\\
			&\preceq \big\Vert\begin{bmatrix}
				\Delta A_i(t)&\Delta B_i(t)
			\end{bmatrix}\big\Vert^2\left|\begin{bmatrix}
				x(t)\\ u(t)
			\end{bmatrix}\right|^2I\\
			&\preceq L^2|t-t^S_i|^2\left|\begin{bmatrix}
				x(t)\\ u(t)
			\end{bmatrix}\right|^2I,
			\end{align*}
			which further implies that
			\begin{equation*}
				W_iW_i^\top = \sum_{t=t^W_i}^{t^S_i-1}w_i(t)w_i^\top(t)\preceq L^2\sum_{k=1}^{T^W}k^2\left|\begin{bmatrix}
					x(t^S_i-k)\\ u(t^S_i-k)
				\end{bmatrix}\right|^2I,
			\end{equation*}
			or equivalently
			\begin{equation}\label{eqn:step_3_in_lemma_1}
				\begin{bmatrix}
					I&W_i
				\end{bmatrix}\begin{bmatrix}
					\Pi&0\\0&-I
				\end{bmatrix}\begin{bmatrix}
					I&W_i
				\end{bmatrix}^\top\succeq 0,
			\end{equation}
			where 
			\begin{equation}\label{def:Pi}
				\Pi:=L^2\sum_{k=1}^{T^W}k^2\left|\begin{bmatrix}
					x(t^S_i-k)\\ u(t^S_i-k)
				\end{bmatrix}\right|^2I.
			\end{equation}
			Although we do not know the exact values of $A_i, B_i$, we can define a set of pairs $(A_i, B_i)$ that are consistent with the collected data $X_i$, $X_i^+$, $U_i$, and ``virtual disturbance'' $W_i$ (which is not measured), \textit{i.e.},
			\begin{align}\label{def:Sigma_i}
				\Sigma_i :=&\ \big\{(A_i,B_i)\in\R^{n\times n}\times\R^{n\times m}: \nonumber\\
				&\ \ \ \exists W_i\mbox{ s.t. }\eqref{sys:LTI_w_disturbance_stack}\mbox{ and }\eqref{eqn:step_3_in_lemma_1} \mbox{ hold.}\big\}
			\end{align}
			Note that $\Sigma_i$ is associated with the state and input data collected over $\calT^W_i$ and such a relationship is indicated by the time window index $i$, for conciseness. This definition is made in the same spirit as the one defined in \cite{HJVW-JE-HLT-MKC:20} to characterize \emph{data informativity}. Interested readers may refer to the explanations therein for further detail.
						We now proceed to estimate the time-varying matrices $\Delta A_i$ and $\Delta B_i$ (the time arguments are omitted for conciseness). Recall Assumption~\ref{ass:Lipschitz} and note that~\eqref{eqn:Lipschitz} yields
						\begin{equation*}
							\begin{bmatrix}
								\Delta A_i&\Delta B_i
							\end{bmatrix}\begin{bmatrix}
								\Delta A_i&\Delta B_i
							\end{bmatrix}^\top\preceq L^2|t-t^S_i|^2I\preceq L^2T^2I
						\end{equation*}
						for any $t\in\calT_i$. This inequality can be equivalently written as
						\begin{equation}\label{error_model}
							\begin{bmatrix}
								I&\Delta A_i&\Delta B_i
							\end{bmatrix}\begin{bmatrix}
								L^2T^2I &0 &0\\
								0& -I &0\\0&0 &-I
							\end{bmatrix}\begin{bmatrix}
								I\\\Delta A_i^\top\\\Delta B_i^\top
							\end{bmatrix}\succeq 0.
						\end{equation}
						Similar to the spirit of~\eqref{def:Sigma_i}, one can define a set of consistent pairs $(\Delta A_i, \Delta B_i)$, that is
						\begin{equation}
							\Sigma^D:=\{(\Delta A_i, \Delta B_i)\in\R^{n\times n}\times\R^{n\times m}:\eqref{error_model} \mbox{ holds.}\}
						\end{equation}
						The set $\Sigma^D$, unlike $\Sigma_i$, is a sheer result of Assumption~\ref{ass:Lipschitz}. Thus, $\Sigma^D$ depends on neither the collected data nor the index $i$.
						
						Since $A(t)=A_i+\Delta A_i(t)$, $B(t)=B_i+\Delta B_i(t)$ for all $t\in\calT_i$, we conclude that $(A(t), B(t))\in\Sigma_i\oplus\Sigma^D$ for all $t\in\calT_i$. Thus in order to solve Problem~\ref{prob:1}, we aim to find a common control gain $K_i$ for the system \emph{uniformly} with respect to the set $\Sigma_i\oplus\Sigma^D$. As such, we arrive at the following formal problem:
						\begin{problem}[Finding the feedback gain]\label{prob:2}
							For the given $\lambda$ as in Assumption~\ref{ass:controllability}, use the collected data $X_i$, $X_i^+$, $U_i$ to find a positive definite matrix $P_i\in\R^{n\times n}$, a matrix $K_i\in\R^{m\times n}$ such that
							\begin{equation}\label{Lyapunov}
								(A+BK_i)^\top P_i(A+BK_i)\preceq \lambda P_i,
							\end{equation}
							for all $(A,B)\in\Sigma_i\oplus\Sigma^D$. 
						\end{problem}
						
							If we take $u=K_ix$, then the function $V_i(x)=x^\top P_i x$ has the property that $V_i(x(t+1))\leq \lambda V_i(x(t))$ for all $t\in\calT_i$, regardless of the values of $A(t), B(t)$. Clearly, $V_i$ will play a large role in the stability analysis, making the resolution of Problem~\ref{prob:2} an important step towards solving Problem~\ref{prob:1}. 
						
						
						The following proposition provides LMI conditions, under which we can solve Problem~\ref{prob:2}.
						\begin{proposition}[An LMI for the feedback gain]\label{thm:1}
							Given a scalar $\lambda\in(0,1)$ and the collected data $X_i$, $X_i^+$, $U_i$ over $\calT^W_i$.
							If there exist scalars $\alpha_1\geq 0$, $\alpha_2\geq 0$, a positive definite matrix $Q_i\in\R^{n\times n}$, and a matrix $L_i\in\R^{m\times n}$ such that
							
							\begin{equation}\label{eqn:S-procedure_LMI}
								\bar M-\alpha_1\bar N_1-\alpha_2\bar N_2\succeq 0,
							\end{equation}
							where $\bar{M}$, $\bar{N}_1$, and $\bar{N}_2$ are as in \eqref{def:bar_M}
							\begin{figure*}[h!]
								\footnotesize
								\vspace*{4pt}
								\hrulefill
								\begin{equation} \label{def:bar_M}
									\bar M:=\!\begin{bmatrix}
										\lambda Q_i\!\!&0&0&0&0&0\\
										0&0&0&0&0&Q_i\\
										0&0&0&0&0&L_i\\
										0&0&0&0&0&Q_i\\
										0&0&0&0&0&L_i\\
										0&Q_i&L_i^\top&Q_i&L_i^\top&Q_i
									\end{bmatrix}\!\!, 
									\bar{N}_1:=\!\begin{bmatrix}
										I &X_i^+\\0 &-X_i\\0 & -U_i\\0&0\\0&0\\0 &0
									\end{bmatrix}\begin{bmatrix}
										\Pi&0\\0&-I
									\end{bmatrix}\begin{bmatrix}
										I &X_i^+\\0 &-X_i\\0 & -U_i\\0&0\\0&0\\0 &0
									\end{bmatrix}^{\!\!\top}\!\!,
									\bar{N}_2:=\!\begin{bmatrix}
										I&0&0\\0&0&0\\0&0&0\\0&I&0\\0&0&I\\0&0&0
									\end{bmatrix}\begin{bmatrix}
										L^2T^2I &0 &0\\
										0& -I &0\\0&0 &-I
									\end{bmatrix}\begin{bmatrix}
										I&0&0\\0&0&0\\0&0&0\\0&I&0\\0&0&I\\0&0&0    
									\end{bmatrix}^{\!\!\top}\!\!.
								\end{equation}
								\normalsize
								\vspace*{4pt}
								\hrulefill
							\end{figure*}
							and $\Pi$ is as defined in \eqref{def:Pi}. Then, the matrices $K_i:=L_iQ_i^{-1}$ and $P_i:=Q_i^{-1}$ solve Problem~\ref{prob:2}.
						\end{proposition}
						The proof of Proposition~\ref{thm:1} is based on methods developed in \cite{HJVW-MKC-JE-HLT:22} and is provided in the Appendix.
						
						\begin{remark}[Proposition~\ref{thm:1} is only sufficient]
							The LMI formulation \eqref{eqn:S-procedure_LMI} is inspired by the S-procedure in \cite[Chapter 2.6]{SB-LEG-EF-VB:94}. However, not all solutions of Problem~\ref{prob:2} can be found by solving this LMI. This is because the Minkowski sum of two sets of matrices defined via quadratic matrix inequalities (\textit{i.e.}, $\Sigma_i$ and $\Sigma^D$) cannot be expressed by another quadratic matrix inequality in general. Conservatism introduced by formulating Problem~\ref{prob:2} into the LMI \eqref{eqn:S-procedure_LMI} can be further investigated in future research.
						\end{remark}

							\section{Stability analysis and algorithmic realization}\label{sec:stability}
							We summarize here that the proposed ODDAC algorithm applies the control law \eqref{true_feedback} to the system \eqref{sys}, where the gain $K_i$ can be computed via solving the LMI in Proposition~\ref{thm:1}. In this section, we will discuss the stability property of the closed-loop system equipped with ODDAC and present an algorithmic realization of the proposed control scheme.
							
							\subsection{Stability analysis}\label{subsec:stability_analysis}
							Note that since the control gain is updated at each time instant $t^S_i$, the closed-loop system can be viewed as a switched system and hence it is stable if there exists a common Lyapunov function (see \cite[Section~2.1]{Liberzon2003b}). In other words, stability is guaranteed when $P_i$'s found by solving Problem~\ref{prob:2} are the same for all $i\in\N$ (including the one given initially). Such a condition, however, is restrictive as it imposes an equality constraint to the subsequent computation of $K_j$ for all $j\geq i$. We therefore adopt an alternative approach to establish stability to allow a different $P_i$ for each control gain update, stated as follows.
							\begin{theorem}[Stability of the closed-loop system]\label{thm:stability}
								Consider the discrete-time LTV system \eqref{sys} equipped with the control law~\eqref{true_feedback}, under Assumption~\ref{ass:Lipschitz} and Assumption~\ref{ass:controllability}. If the LMI \eqref{eqn:S-procedure_LMI} is feasible for all $i\in\Np$ and the matrices $P_i:=Q_i^{-1}$  satisfy 
								\begin{align}
									\sigma_1 I\preceq P_i\preceq \sigma_2 I,\label{sandwich}\\
									P_{i+1}\preceq \left(\frac{\hat\lambda}{\lambda}\right)^{T}P_i,\label{condition_gain_at_switch}
								\end{align}
								for all $i\in\N$ and some $\hat\lambda \in[\lambda,1)$, $\sigma_1>0$, $\sigma_2>0$,
								then, the system is pGES.
								In other words, the solutions of the closed-loop system satisfy that
							\begin{equation}\label{final_practical_stability}
								|x(t)|\leq \frac{\sigma_2}{\sqrt{\sigma_1}}\hat\lambda^{\frac{t}{2}}|x(0)|+\sqrt{\frac{\sigma_2}{\sigma_1}}\left(1-\sqrt{\hat\lambda}\right)^{-1}\left(\frac{\hat\lambda}{\lambda}\right)^{\frac{T}{2}}\bar B\bar v,
							\end{equation}
							where $\bar v$ is the uniform bound of $v$, \textit{i.e.}, $|v(t)|\leq \bar v$ for all $t\in\cup_{i\in\Np}\calT^W_i$.
						\end{theorem}
						
						The proof of Theorem~\ref{thm:stability} exploits standard arguments of multiple Lyapunov function approach that has been extensively used in the literature (see \textit{e.g.}, \cite{HL-PJA:09}) and is presented in the Appendix.
						
						We now make some comments to facilitate the understanding of Theorem~\ref{thm:stability}. Equation~\eqref{sandwich} is essentially the ``sandwich" condition typically required for the multiple Lyapunov function approach. Meanwhile, if there exists $\mu\geq 1$ such that $P_{i+1}\preceq \mu P_i$ for all $i\in\N$, it is sufficient to require
						\begin{equation}\label{DT}
							T>-\frac{\ln\mu}{\ln\lambda},
						\end{equation}
						in order for \eqref{condition_gain_at_switch} to hold, in which case $\hat\lambda\in[\lambda\mu^{\frac{1}{T}},1)$. Equation~\eqref{DT} imposes a dwell time condition on the switching \cite{ASM:93}. This can be intuitively understood as follows: a switched system is stable if it is stable in each mode, and the switching is sufficiently slow. 
						With this controller, we observe in \eqref{final_practical_stability} that the solutions can converge to a ball of arbitrarily small radius, by making $\bar v$ sufficiently small. In theory, one can establish global asymptotic stability for the closed-loop system by selecting a exciting signal $v(t)$ that depends on $|x(t)|$. This is however not practically useful as numerical issues arise in the solution of the LMIs if $|v|$ is too small.
								\subsection{The control algorithm}
								
								Recall that Proposition~\ref{thm:1} gave an LMI approach to solve Problem~\ref{prob:2}, which is not directly formulated in terms of $P_i, K_i$. Yet, our stability results rely on some additional conditions imposed on $P_i$; namely, the conditions~\eqref{sandwich} and \eqref{condition_gain_at_switch}. We remark here that with given values of $\hat\lambda$, $\sigma_1$,and $\sigma_2$, these two conditions can be easily encoded as additional LMIs with respect to variables $Q_i=P_i^{-1}$. To do this, note that \eqref{sandwich} is equivalent to
								\begin{equation}\label{sandwich_LMI}
									\sigma_2^{-1}I\preceq Q_i\preceq \sigma_1^{-1}I.
								\end{equation}
								Meanwhile, multiply \eqref{condition_gain_at_switch} by $Q_i$ on both sides, we get
								\begin{equation*}
									Q_iP_{i+1}Q_i\preceq \left(\frac{\hat\lambda}{\lambda}\right)^{T}Q_i,
								\end{equation*}
								which, by exploiting Schur complement, can be equivalently written as
								\begin{equation}\label{condition_gain_at_switch_LMI}
									\begin{bmatrix}
										\hat\lambda ^T Q_i & Q_i\\Q_i &\lambda ^{-T}Q_{i+1}
									\end{bmatrix}\succeq 0.
								\end{equation}

								Based on the timing mechanism sketched in Section~\ref{subsec:sketch}, we can now summarize the control algorithm as in Algorithm~\ref{alg:OADDC}.
								\begin{algorithm}[htb!]
									\caption{Online Data-Driven Adaptive Control}\label{alg:OADDC}
									\algorithmicrequire $K_0$, $P_0$, $\lambda$, $T$, $T^W$
									\begin{algorithmic}[1]
										\State $i\gets0$, $t\gets0$
										\State Set $\hat\lambda\in(\lambda,1)$, $\sigma_1>0$, $\sigma_2>0$, and $\bar v>0$.
										\State $u(0)\gets K_0x(0)$     
										\State $Q_0\gets P_0^{-1}$
										\State Set $X_i,X_i^+,U_i$ as empty matrices      
										\While{system is running}
										\State $i_{\rm new}=\lfloor\frac{t}{T}\rfloor$
										\If{$t-i_{\rm new}T\geq T-T^W$}
										\Comment{{\color{comment} When $t\in\calT_{i+1}^W$}}
										\State Append $x(t),x(t+1),u(t)$ to $X_i,X_i^+,U_i$
										\Comment{{\color{comment}
												\Statex\hskip\algorithmicindent\hskip\algorithmicindent
												Collect the new data}}
										\State $u(t)\gets K_ix(t)+v(t)$
										\Comment{{\color{comment} Update the control
												\Statex\hskip\algorithmicindent\hskip\algorithmicindent
												input with the exciting signal $v$}}
										\Else
										\Comment{{\color{comment} When $t\in\calT_i\backslash\calT_{i+1}^W$}}
										\If{$i_{\rm new}\neq i$}
										\Comment{{\color{comment} When $t=t_{i_{new}}^S$}}
										\State $i\gets i_{\rm new}$
										\State Solve the LMIs \eqref{eqn:S-procedure_LMI}, \eqref{sandwich_LMI}, and \eqref{condition_gain_at_switch_LMI}
										\Statex\hskip\algorithmicindent\hskip\algorithmicindent\hskip\algorithmicindent
										for $Q_i, L_i, \alpha_1, \alpha_2$
										\State $K_i\gets L_iQ_i^{-1}$
										\Comment{{\color{comment} Update the control gain}}
										\State Reset $X_i,X_i^+,U_i$ as empty matrices
										\EndIf
										\State $u(t)\gets K_ix(t)$
										\Comment{{\color{comment} Update the control input 
												\Statex\hskip\algorithmicindent\hskip\algorithmicindent
												without the exciting signal $v$}}
										\EndIf
										\State $t\gets t+1$
										\EndWhile
									\end{algorithmic}
								\end{algorithm}
								In the pseudo-code of the algorithm, Lines~1--5 are the initialization procedures. For $t\in\calT_i\backslash\calT_{i+1}^W$, which leads to the branch starting from Line~11, the control $u(t)=K_ix(t)$ is applied to the system, where the gain $K_i$ is either initialized at the beginning of the control process, or computed in the last while-loop. This is consistent with the first line in \eqref{true_feedback}. When the time reaches $t^W_{i+1}$, which triggers the condition in Line~8, the controller activates the exciting signal and starts to collect the data. As the time reaches $t^S_i$, which triggers the condition in Line~12, a new feedback gain is computed, by solving the LMIs \eqref{eqn:S-procedure_LMI}, \eqref{sandwich_LMI} and \eqref{condition_gain_at_switch_LMI}. Meanwhile, the index $i$ is updated, indicating that a new period has started, and the data buffers $X_i, X^+_i, U_i$ are reset. Thanks to Theorem~\ref{thm:stability}, we have the following result regarding the functionality of Algorithm~\ref{alg:OADDC}.
								
								\begin{corollary}[Functionality of the ODDAC algorithm]
									Consider the discrete-time LTV system \eqref{sys} under Assumption~\ref{ass:Lipschitz} and Assumption~\ref{ass:controllability}. Let the ODDAC Algorithm~\ref{alg:OADDC} be applied to the system and assume that Line~14 of the algorithm is always feasible. Then the closed-loop system is pGES, in the sense that~\eqref{final_practical_stability} holds on all solutions.
								\end{corollary}
								
								\section{Simulation}\label{sec:simulation}
								In this section, we present a numerical example to which our controller is applied and compare the performance of the proposed controller to that of some control schemes in the literature. 
								
\begin{figure}
									\tikzset{every picture/.style={scale=0.8}}
									\centering
									\input{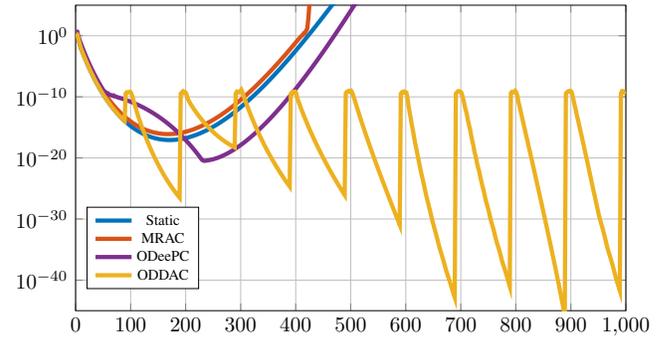}
									\caption{Semi-logarithmic time histories of $|x(t)|$ of the LTV system in the four control schemes.}\label{fig:comparison}
								\end{figure}

\begin{figure}
									\tikzset{every picture/.style={scale=0.8}}
									\centering
									\input{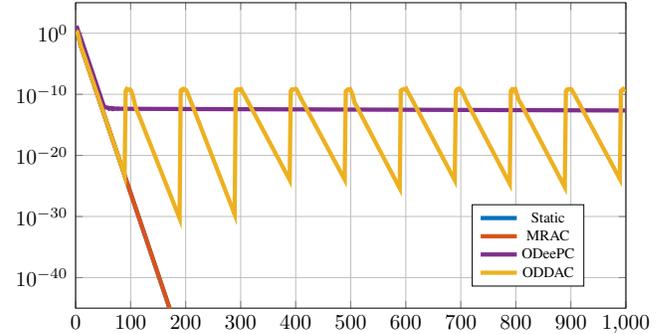}
									\caption{Semi-logarithmic time histories of $|x(t)|$ of the LTI system in the four control schemes.}\label{fig:comparison_2}
								\end{figure}

								Consider an LTV system in the form of~\eqref{sys} with $n=5, m=2$. The time-varying $A(t)$ and $ B(t)$ are generated by the (element-wise) cubic interpolation of the three pairs of matrices:
								
								\begin{align*}
									A(0)\!&=\!\begin{bmatrix}\begin{smallmatrix}
											-0.5   &-0.4   & 0.1   &-0.8    &-0.2\\
											-0.5   &-0.1   & 0.2   & 0.7    &   0\\
											-0.4   &-0.9   & 0.6   &-0.3    & 0.4\\
											0.2   &-0.3   &-1.2   &   0    &-0.1\\
											-0.6   & 0.8   &-0.5   &-0.1    &-0.1\\
									\end{smallmatrix}\end{bmatrix}\!, 	B(0)\!=\!\begin{bmatrix}\begin{smallmatrix}
											-1.4    & 2.2\\
											0.9    & 1.4\\
											2.7    & 0.5\\
											-0.7    & 1.5\\
											0.6    &-1.9
									\end{smallmatrix}\end{bmatrix}\!,\\
									A(500)\!&=\!\begin{bmatrix}\begin{smallmatrix}
											-0.5   &-0.7   & 0.3   &-0.6    &   0\\
											0   &   0   &   0   & 0.8    & 0.4\\
											-0.7   &-1.0   & 0.7   & 0.1    & 0.2\\
											-0.2   &-0.2   &-1.1   & 0.3    & 0.3\\
											-0.9   & 0.7   &-0.9   & 0.5    & 0.4		
									\end{smallmatrix}\end{bmatrix}\!,B(500)\!=\!\begin{bmatrix}\begin{smallmatrix}
											-1.5    & 2.4\\
											0.9    & 1.3\\
											2.9    & 0.7\\
											-0.7    & 1.5\\
											0.4    &-1.9
									\end{smallmatrix}\end{bmatrix}\!,\\
									A(1000)\!&=\!\begin{bmatrix}\begin{smallmatrix}
											0   &-0.6    &-0.2    &-0.7    & 0.5\\
											0   & 0.1    & 0.4    & 1.1    & 0.7\\
											-1.4   &-0.9    & 0.5    & 0.5    & 0.5\\
											-0.2   &-0.2    &-1.5    &-0.3    & 0.5\\
											-0.9   & 0.5    &-0.6    & 0.7    & 0.5\\
									\end{smallmatrix}\end{bmatrix}\!,B(1000)\!=\!\begin{bmatrix}\begin{smallmatrix}
											-1.4    & 2.4\\
											0.9    & 1.5\\
											3.0    & 0.6\\
											-0.8    & 1.5\\
											0.5    &-1.9
									\end{smallmatrix}\end{bmatrix}\!.
								\end{align*}
								One can numerically verify that Assumption~\ref{ass:Lipschitz} holds with the Lipschitz constant $L=0.0037$. Assumption~\ref{ass:controllability} also holds with initial values of $K_i, Q_i$ given by
								\begin{align*}
									K_0&=\begin{bmatrix}
										0.13    &0.26   &-0.25   &0.04   &-0.13\\
										0.08    &0.28   & 0.13   &0.05   & 0.01 			
									\end{bmatrix},\\
									Q_0&=\begin{bmatrix}
										0.75 &-0.13 & 0.03 &-0.26 &-0.08\\
										-0.13 & 0.88 &-0.08 &-0.12 & 0.36\\
										0.03 &-0.08 & 0.21 & 0.01 &-0.01\\
										-0.26 &-0.12 & 0.01 & 0.43 & 0.14\\
										-0.08 & 0.36 &-0.01 & 0.14 & 1.13
									\end{bmatrix}.
								\end{align*}
								Let the initial state be $x(0)=\begin{bmatrix}
									1&1&1&1&1
								\end{bmatrix}^\top$. We consider four different control methods to stabilize the system: 1) static state feedback control $u=K_0x$, 2) a discrete-time version of the model reference adaptive control (MRAC) in \cite[Section~5.2.6]{ioannou2006adaptive} with normalization \cite[Section~4.3]{ioannou1996adaptive}, 3) online data-enabled predictive control (ODeePC) as recently proposed in \cite{Baros2022}, and 4) the proposed ODDAC method. Assume that the matrix trajectories $A(t), B(t)$ are unknown to all four controllers. In particular, the parameters of ODDAC are selected such that $T=100$, $T^W=10$, $\lambda=0.9$, $\hat\lambda=0.91$, $\sigma_1=0.001$, $\sigma_2=1000$, $\bar v=10^{-10}$. 						
								The semi-logarithmic time histories of $|x(t)|$ are shown in Fig.~\ref{fig:comparison}. 
								We also apply the four control schemes with the same parameters to the LTI system (\textit{i.e.}, $(A(t),B(t))=(A(0),B(0))$ for all $t\in\N$). The semi-logarithmic time histories of $|x(t)|$ of the LTI system are plotted in Fig.~\ref{fig:comparison_2}, where it is seen that both static state feedback control and MRAC make the closed-loop system exponentially stable, and both ODeePC and ODDAC make the closed-loop system pGES.
								
								From Fig.~\ref{fig:comparison}, one can observe that all the controllers can guarantee that the solutions of the LTV system evolve close to the origin over a short period of time. Nevertheless, only ODDAC can maintain this for the entire simulated time interval. For all other methods, the solutions of the closed-loop system diverge after some time.
								Note that the ``sawtooth'' pattern of $\log|x(t)|$ in the ODDAC time history has a time period of $100$ time units. This is caused by the exciting signal $v(t)$, which drives $|x(t)|$ to an acceptable amplitude (depends on the magnitude of $\bar v$) to collect the data for computing the new feedback gain. This is the ``price'' paid to guarantee practical stability for an arbitrarily long time. The simulation results are hence consistent with the practical stability property established in the paper and reveal the advantage of the proposed control algorithm in time-varying systems.
								
								\section{Conclusion}\label{sec:conclusion}
								This paper has proposed and discussed a novel control algorithm for LTV systems: ODDAC. Based on the data collected over the preceding short time window, the algorithm periodically updates a feedback gain, aiming to stabilize the system up to the time of the next update. In order to guarantee pGES of the closed-loop system equipped with ODDAC, we assume that the LMIs for finding the feedback gains always have solutions. Although feasibility of these LMIs is expected to be true when the system varies sufficiently slowly, this assumption needs to be further investigated in detail. In the meantime, other future research includes taking measurement error into account, modification of the algorithm for the case when the system dynamics is partially known, and a more involved adaptive controller design via dynamic feedback instead of switched static feedback.

								\appendix
								
								\textit{Proof of Proposition~\ref{thm:1}.} First of all, by Schur complement, \eqref{eqn:S-procedure_LMI} is equivalent to
								\begin{equation}\label{eqn:S-procedure}
									M-\alpha_1 N_1-\alpha_2N_2\succeq 0,
								\end{equation}
								where
								\begin{align}\label{def:M}
									M:&=\begin{bmatrix}
										\lambda Q_i&0&\cdots&0\\
										0&0&\cdots & 0\\
										\vdots&\vdots &\ddots&\vdots\\
										0& 0 &\cdots &0
									\end{bmatrix}-\begin{bmatrix}
										0\\Q_i\\L_i\\Q_i\\L_i
									\end{bmatrix}Q_i^{-1}\begin{bmatrix}
										0\\Q_i\\L_i\\Q_i\\L_i
									\end{bmatrix}^\top,\\
									N_1:&=\begin{bmatrix}
										I &X_i^+\\0 &-X_i\\0 & -U_i\\0&0\\0&0
									\end{bmatrix}\begin{bmatrix}
										\Pi&0\\0&-I
									\end{bmatrix}\begin{bmatrix}
										I &X_i^+\\0 &-X_i\\0 & -U_i\\0&0\\0&0
									\end{bmatrix}^\top,\label{def:N1}\\
									N_2:&=\begin{bmatrix}
										I&0&0\\0&0&0\\0&0&0\\0&I&0\\0&0&I
									\end{bmatrix}\begin{bmatrix}
										L^2T^2I &0 &0\\
										0& -I &0\\0&0 &-I
									\end{bmatrix}\begin{bmatrix}
										I&0&0\\0&0&0\\0&0&0\\0&I&0\\0&0&I
									\end{bmatrix}^\top.\label{def:N2}
								\end{align}
								
								Define 
								\begin{equation*}
									\Theta_i:=\begin{bmatrix}I&A_i&B_i&\Delta A_i&\Delta B_i\end{bmatrix}\in \R^{n\times (3n+2m)}.
								\end{equation*}
								It follows from \eqref{sys:LTI_w_disturbance_stack} that
								\begin{equation}
									\begin{bmatrix}
										I& W_i
									\end{bmatrix}=\Theta_i\begin{bmatrix}
										I &X_i^+\\0 &-X_i\\0 & -U_i\\0&0\\0&0
									\end{bmatrix}.
								\end{equation}
								Hence \eqref{eqn:step_3_in_lemma_1} is equivalent to
								\begin{equation}\label{part_N1}
									\Theta_i N_1 \Theta_i^\top\succeq 0.
								\end{equation}
									Meanwhile,
									\begin{equation}
										\begin{bmatrix}
											I&\Delta A_i&\Delta B_i
										\end{bmatrix}=\Theta_i\begin{bmatrix}
											I&0&0\\0&0&0\\0&0&0\\0&I&0\\0&0&I
										\end{bmatrix}.
									\end{equation}
									Hence, \eqref{error_model} can be equivalently written as
									\begin{equation}\label{part_N2}
										\Theta_i N_2 \Theta_i^\top\succeq 0.
									\end{equation}
									Finally, we show that the condition \eqref{Lyapunov}
									can also be converted to a similar quadratic form. To this end, we first define $Q_i:=P_i^{-1}$. Then it can be shown by Schur complement that \eqref{Lyapunov} is equivalent to the inequality
									\begin{multline}\label{Lyapunov2}
										(A_i+\Delta A_i +(B_i+\Delta B_i)K_i) Q_i(A_i+\Delta A_i +(B_i+\Delta B_i)K_i)^\top\\\preceq \lambda Q_i.
									\end{multline}
									Next, since 
									\begin{equation*}
										A_i+\Delta A_i +(B_i+\Delta B_i)K_i=\Theta_i \begin{bmatrix}
											0&I&K_i^\top&I&K_i^\top
										\end{bmatrix}^\top,
									\end{equation*}
									the inequality \eqref{Lyapunov2} is equivalent to
									\begin{equation*}\Theta_i\begin{bmatrix}0\\I\\K_i\\I\\K_i\end{bmatrix}Q_i\begin{bmatrix}0\\I\\K_i\\I\\K_i\end{bmatrix}^\top \Theta_i^\top\preceq \lambda Q_i.
									\end{equation*}
									Plug in $K_i=L_iQ_i^{-1}$, we conclude that with $M$ as in \eqref{def:M}, 
									\begin{equation}\label{part_M}
										\Theta_i M\Theta_i^\top \succeq 0.
									\end{equation}
									Now, with all the aforementioned identities, Problem~\ref{prob:2} is equivalent to~\eqref{part_M} for all $\Theta_i\in \R^{n\times (3n+2m)}$ such that \eqref{part_N1} and \eqref{part_N2} hold.
									Clearly, when inequality \eqref{eqn:S-procedure} holds for some $\alpha_1,\alpha_2\geq 0$, we have
									\begin{multline*}
										\Theta_iM\Theta_i^\top\succeq \Theta_i(\alpha_1N_1+\alpha_2N_2)\Theta_i^\top\\
										= \alpha_1\Theta_iN_1\Theta_i^\top+\alpha_2\Theta_iN_2\Theta_i^\top\succeq 0,
									\end{multline*}
									which proves Proposition~\ref{thm:1}.\hfill$\square$

									\textit{Proof of Theorem~\ref{thm:stability}.} Define 
									\begin{equation}
										q(t):=\left(\frac{\hat\lambda}{\lambda}\right)^{\frac{t-iT}{2}}|P_i^{\frac{1}{2}}x(t)|,
									\end{equation}
									for all $t\in\calT_i, i\in\N$.
									Note that $q$ only depends on $t$ (not on $i$ because $i=\lfloor\frac{t}{T}\rfloor$). 
									We first claim that when the conditions in Theorem~\ref{thm:stability} hold, then
									\begin{align}
										\sqrt{\sigma_1}|x(t)|&\leq q(t)\leq \left(\frac{\hat\lambda}{\lambda}\right)^{\frac{T-1}{2}}\sqrt{\sigma_2}|x(t)|,\label{sandwich_V}\\
										q(t+1)&\leq \sqrt{\hat\lambda} q(t)+\left(\frac{\hat\lambda}{\lambda}\right)^{\frac{T}{2}}\sqrt{\sigma_2}\bar B\bar v\label{consecutive_V}
									\end{align}
									for all $t\in\N$.
									
									The claim \eqref{sandwich_V} is not difficult to show, by noting the fact that $0\leq t-iT\leq T-1$,
									for all $t\in\calT_i$,
									and the fact that $|P_i^{\frac{1}{2}}x|^2=x^\top P_i x$. Hence it follows from \eqref{sandwich} that $\sigma_1|x|^2\leq |P_i^{\frac{1}{2}}x|^2\leq \sigma_2|x|^2$. 
									
									We then prove the claim \eqref{consecutive_V}. Define
									\begin{equation*}
										\tilde v(t):=\begin{cases}
											0, &t\in\calT_i\backslash\calT^W_{i+1},\ i\in\N,\\
											v(t), &t\in\calT^W_{i+1},\ i\in\N.
										\end{cases}
									\end{equation*}
									Then $u(t)=K_ix(t)+\tilde v(t)$ for all $t\in\calT_i, i\in\N$. With this notion, it follows from \eqref{Lyapunov} and triangle inequality that
									\begin{align*}
										&\hspace{-.5cm}\sqrt{x^\top(t+1)P_ix(t+1)}\\
										&=|P_i^{\frac{1}{2}}(A(t)x(t)+B(t)u(t))|\\
										&=\big|P_i^{\frac{1}{2}}\big((A(t)+B(t)K_i)x(t)+B(t)\tilde v(t)\big)\big|\\
										&\leq |P_i^{\frac{1}{2}}(A(t)+B(t)K_i)x(t)|+|P_i^{\frac{1}{2}}B(t)\tilde v(t)|\\
										&=\sqrt{x^\top(t)(A(t)+B(t)K_i)^\top P_i(A(t)+B(t)K_i)x(t)}\\
										&\quad+|P_i^{\frac{1}{2}}B(t)\tilde v(t)|\\
										&\leq \sqrt{\lambda x^\top(t)P_ix(t)}+\sqrt{\sigma_2}\bar B\bar v.
									\end{align*}
									In other words, we have
									\begin{equation}\label{norm_difference}
										|P_i^{\frac{1}{2}}x(t+1)|\leq\sqrt{\lambda}|P_i^{\frac{1}{2}}x(t)|+\sqrt{\sigma_2}\bar B\bar v,
									\end{equation}
									for all $t\in\calT_i, i\in\N$.
									We now consider two cases. In the first case, for any $t\in\calT_i\backslash\{t^S_i+T-1\}, i\in\N$, we have $t+1\in\calT_i$ as well. 
									Hence it follows from \eqref{norm_difference} that
									\begin{align*}
										q(t+1)&=\left(\frac{\hat\lambda}{\lambda}\right)^{\frac{t+1-iT}{2}}|P_i^{\frac{1}{2}}x(t+1)|\\
										&\leq \left(\frac{\hat\lambda}{\lambda}\right)^{\frac{t+1-iT}{2}}\left(\sqrt{\lambda}|P_i^{\frac{1}{2}}x(t)|+\sqrt{\sigma_2}\bar B\bar v\right)\\
										&\leq \sqrt{\hat\lambda}\left(\frac{\hat\lambda}{\lambda}\right)^{\frac{t-iT}{2}}|P_i^{\frac{1}{2}}x(t)|+\left(\frac{\hat\lambda}{\lambda}\right)^{\frac{T-1}{2}}\sqrt{\sigma_2}\bar B\bar v\\
										&=\sqrt{\hat\lambda}q(t)+\left(\frac{\hat\lambda}{\lambda}\right)^{\frac{T-1}{2}}\sqrt{\sigma_2}\bar B\bar v.
									\end{align*}
									In the second case, for any $t=t^S_i+T-1, i\in\N$, we have $t+1=t^S_{i+1}\in\calT_{i+1}$. Therefore, $t+1=(i+1)T$ and it follows from \eqref{condition_gain_at_switch} and \eqref{norm_difference} that
									\begin{align*}
										q(t+1)&=|P_{i+1}^{\frac{1}{2}}x(t+1)|\\
										&\leq \left(\frac{\hat\lambda}{\lambda}\right)^{\frac{T}{2}}|P_{i}^{\frac{1}{2}}x(t+1)|\\
										&\leq \left(\frac{\hat\lambda}{\lambda}\right)^{\frac{T}{2}}\left(\sqrt{\lambda}|P_i^{\frac{1}{2}}x(t)|+\sqrt{\sigma_2}\bar B\bar v\right)\\
										&=\sqrt{\hat\lambda}\left(\frac{\hat\lambda}{\lambda}\right)^{\frac{T-1}{2}}|P_i^{\frac{1}{2}}x(t)|+\left(\frac{\hat\lambda}{\lambda}\right)^{\frac{T}{2}}\sqrt{\sigma_2}\bar B\bar v\\
										&=\sqrt{\hat\lambda}q(t)+\left(\frac{\hat\lambda}{\lambda}\right)^{\frac{T}{2}}\sqrt{\sigma_2}\bar B\bar v.
									\end{align*}
									Hence in both cases, we have proven that \eqref{consecutive_V} holds.
									
									Now by iteratively applying the inequality \eqref{consecutive_V}, we conclude that
									\begin{align*}
										q(t)&\leq \hat\lambda^{\frac{t}{2}}q(0)+\left(\sum_{i=0}^{t-1}\hat\lambda^{\frac{i}{2}}\right)\left(\frac{\hat\lambda}{\lambda}\right)^{\frac{T}{2}}\sqrt{\sigma_2}\bar B\bar v\\
										&=\hat\lambda^{\frac{t}{2}}q(0)+\left(\frac{1-\hat\lambda^{\frac{t}{2}}}{1-\hat\lambda^{\frac{1}{2}}}\right)\left(\frac{\hat\lambda}{\lambda}\right)^{\frac{T}{2}}\sqrt{\sigma_2}\bar B\bar v\\
										&\leq \hat\lambda^{\frac{t}{2}}q(0)+\left(1-\sqrt{\hat\lambda}\right)^{-1}\left(\frac{\hat\lambda}{\lambda}\right)^{\frac{T}{2}}\sqrt{\sigma_2}\bar B\bar v.
									\end{align*}
									Finally, recall the definition of $q$. Using~\eqref{sandwich_V}, we conclude that~\eqref{final_practical_stability} holds. \hfill$\square$									
									
									\bibliographystyle{IEEEtran}
									

\begin{thebibliography}{10}
										\providecommand{\url}[1]{#1}
										\csname url@rmstyle\endcsname
										\providecommand{\newblock}{\relax}
										\providecommand{\bibinfo}[2]{#2}
										\providecommand\BIBentrySTDinterwordspacing{\spaceskip=0pt\relax}
										\providecommand\BIBentryALTinterwordstretchfactor{4}
										\providecommand\BIBentryALTinterwordspacing{\spaceskip=\fontdimen2\font plus
											\BIBentryALTinterwordstretchfactor\fontdimen3\font minus
											\fontdimen4\font\relax}
										\providecommand\BIBforeignlanguage[2]{{%
												\expandafter\ifx\csname l@#1\endcsname\relax
												\typeout{** WARNING: IEEEtran.bst: No hyphenation pattern has been}%
												\typeout{** loaded for the language `#1'. Using the pattern for}%
												\typeout{** the default language instead.}%
												\else
												\language=\csname l@#1\endcsname
												\fi
												#2}}
										
										\bibitem{JCW-PR-IM-BLMDM:05}
										J.~C. Willems, P.~Rapisarda, I.~Markovsky, and B.~L.~M. {De Moor}, ``A note on
										persistency of excitation,'' \emph{Systems \& Control Letters}, vol.~54,
										no.~4, pp. 325--329, 2005.
										
										\bibitem{Coulson2019}
										J.~Coulson, J.~Lygeros, and F.~D\"orfler, ``Data-enabled predictive control: In
										the shallows of the {DeePC},'' in \emph{18th European Control Conference
											(ECC)}, 2019, pp. 307--312.
										
										\bibitem{CDP-PT:19}
										C.~{De Persis} and P.~Tesi, ``Formulas for data-driven control: Stabilization,
										optimality and robustness,'' \emph{IEEE Transactions on Automatic Control},
										vol.~65, no.~3, pp. 909--924, 2019.
										
										\bibitem{Breschi2021}
										V.~Breschi, C.~D. Persis, S.~Formentin, and P.~Tesi, ``Direct data-driven
										model-reference control with Lyapunov stability guarantees,'' in \emph{2021
											60th IEEE Conference on Decision and Control (CDC)}, 2021, pp. 1456--1461.
										
										\bibitem{HJVW-JE-HLT-MKC:20}
										H.~J. van Waarde, J.~Eising, H.~L. Trentelman, and M.~K. Camlibel, ``Data
										informativity: a new perspective on data-driven analysis and control,''
										\emph{IEEE Transactions on Automatic Control}, vol.~65, no.~11, pp.
										4753--4768, 2020.
										
										\bibitem{MR-CDP-PT:22}
										M.~Rotulo, C.~{De Persis}, and P.~Tesi, ``Online learning of data-driven
										controllers for unknown switched linear systems,'' \emph{Automatica}, vol.
										145, p. 110519, 2022.
										
										\bibitem{Eising2022}
										J.~Eising, S.~Liu, S.~Mart\'inez, and J.~Cort\'es, ``Using data informativity
										for online stabilization of unknown switched linear systems,'' in \emph{2022
											IEEE 61st Conference on Decision and Control (CDC)}, 2022, pp. 8--13.
										
										\bibitem{RuedaEscobedo2022}
										J.~G. Rueda-Escobedo, E.~Fridman, and J.~Schiffer, ``Data-driven control for
										linear discrete-time delay systems,'' \emph{IEEE Transactions on Automatic
											Control}, vol.~67, no.~7, pp. 3321--3336, 2022.
										
										\bibitem{TD-MS:20_2}
										T.~Dai and M.~Sznaier, ``A semi-algebraic optimization approach to data-driven
										control of continuous-time nonlinear systems,'' \emph{IEEE Control Systems
											Letters}, vol.~5, no.~2, pp. 487--492, 2020.
										
										\bibitem{MG-CDP-PT:21}
										M.~Guo, C.~D. Persis, and P.~Tesi, ``Data-driven stabilization of nonlinear
										polynomial systems with noisy data,'' \emph{IEEE Transactions on Automatic
											Control}, 2021.
										
										\bibitem{Pang2018}
										B.~Pang, T.~Bian, and Z.-P. Jiang, ``Data-driven finite-horizon optimal control
										for linear time-varying discrete-time systems,'' in \emph{2018 IEEE
											Conference on Decision and Control (CDC)}, 2018, pp. 861--866.
										
										\bibitem{Nortmann2020}
										B.~Nortmann and T.~Mylvaganam, ``Data-driven control of linear time-varying
										systems,'' in \emph{2020 59th IEEE Conference on Decision and Control (CDC)},
										2020, pp. 3939--3944.
										
										\bibitem{Verhoek2021}
										C.~Verhoek, R.~Tóth, S.~Haesaert, and A.~Koch, ``Fundamental lemma for
										data-driven analysis of linear parameter-varying systems,'' in \emph{2021
											60th IEEE Conference on Decision and Control (CDC)}, 2021, pp. 5040--5046.
										
										\bibitem{Baros2022}
										S.~Baros, C.-Y. Chang, G.~E. Col\'on-Reyes, and A.~Bernstein, ``Online
										data-enabled predictive control,'' \emph{Automatica}, vol. 138, p. 109926,
										2022.
										
										\bibitem{JM-MS-22}
										J.~Miller and M.~Sznaier, ``Data-driven gain scheduling control of linear
										parameter-varying systems using quadratic matrix inequalities,'' \emph{arXiv
											preprint arxiv:2209.06251}, 2022.
										
										\bibitem{HJVW:21}
										H.~J. van Waarde, ``Beyond persistent excitation: Online experiment design for
										data-driven modeling and control,'' \emph{IEEE Control Systems Letters},
										vol.~6, pp. 319--324, 2022.
										
										\bibitem{chen2020adaptive}
										K.~Chen and A.~Astolfi, ``Adaptive control for systems with time-varying
										parameters,'' \emph{IEEE Transactions on Automatic Control}, vol.~66, no.~5,
										pp. 1986--2001, 2021.
										
										\bibitem{HJVW-MKC-JE-HLT:22}
										H.~J. van Waarde, M.~K. Camlibel, J.~Eising, and H.~L. Trentelman, ``Quadratic
										matrix inequalities with applications to data-based control,'' \emph{arXiv
											preprint arXiv:2203.12959}, 2022.
										
										\bibitem{SB-LEG-EF-VB:94}
										S.~Boyd, L.~E. Ghaoui, E.~Feron, and V.~Balakrishnan, \emph{Linear Matrix
											Inequalities in System and Control Theory}, ser. {Studies in Applied
											Mathematics}.\hskip 1em plus 0.5em minus 0.4em\relax Philadelphia,
										Pennsylvania: {SIAM}, 1994, vol.~15.
										
										\bibitem{Liberzon2003b}
										D.~Liberzon, \emph{Switching in Systems and Control}.\hskip 1em plus 0.5em
										minus 0.4em\relax Boston, MA: Birkh{\"a}user, 2003.
										
										\bibitem{HL-PJA:09}
										H.~Lin and P.~J. Antsaklis, ``Stability and stabilizability of switched linear
										systems: A survey of recent results,'' \emph{IEEE Transactions on Automatic
											Control}, vol.~54, no.~2, pp. 308--322, 2009.
										
										\bibitem{ASM:93}
										A.~S. {Morse}, ``Dwell-time switching,'' in \emph{{E}uropean {C}ontrol
											{C}onference}, Groningen, The Netherlands, 1993, pp. 176--181.
										
										\bibitem{ioannou2006adaptive}
										P.~A. Ioannou and B.~Fidan, \emph{Adaptive control tutorial}.\hskip 1em plus 0.5em
										minus 0.4em\relax SIAM, 2006.
										
										\bibitem{ioannou1996adaptive}
										P.~A. Ioannou and J.~Sun, \emph{Robust Adaptive Control}.\hskip 1em plus 0.5em
										minus 0.4em\relax Upper Saddle River, NJ: Prentice-Hall, 1996.
										
									\end{thebibliography}

								\end{document}